\documentclass[12pt,a4]{article} 
 
\def \ni {\noindent}
\def \vs  {\vskip5mm}
 
\usepackage{amsmath,amssymb} 
 
\begin{document} 
\begin{titlepage} 
 
\title{Massive neutrinos,  massless neutrinos, and $so(4,2)$ invariance.} 
 
\author{A.J. Bracken
\\
Centre for Mathematical Physics\\
Department of Mathematics\\
University of Queensland\\Brisbane 4072\\Queensland\\Australia\\e-mail:ajb@maths.uq.edu.au} 
 
\date{} 
\maketitle 
 
\begin{abstract} 
Dirac's equation for a massless particle is conformal invariant, and accordingly has an $so(4,2)$
invariance algebra. It is known that although 
Dirac's equation for a massive spin 1/2 particle is not conformal invariant, it too
has an $so(4,2)$ invariance
algebra.  
It is shown here that the algebra of operators associated with
a 4-component massless particle, or two flavors of  2-component massless particles,
can be deformed into the algebra of
operators associated with a spin 1/2 particle with positive rest mass.  
It is speculated that this may be exploited to describe massless neutrino mixing.  
\end{abstract} 
 
\end{titlepage}

\section{Introduction}

\vs\ni
In the 1970s and 1980s,
Asim Barut's name became synonymous
with the applications to quantum mechanics of the Lie algebra $so(4,2)$, whether as a space-time
(conformal) symmetry algebra \cite{barutbrittin},
as a spectrum generating algebra in generalised Coulomb problems \cite{canterbury}, or as
an algebra associated with Dirac's equation for the electron
\cite{barutbracken}.
I would like to revisit from that period an idea involving $so(4,2)$ 
that may have renewed relevance
because of the subsequent discovery of several neutrino types, some of which
may have small rest
masses.  
\vs\ni
It is well known that 
the massless Dirac equation is not
only Poincar\'e invariant but also $so(4,2)$ (conformal) invariant \cite{macksalam}.
More surprising is that the massive Dirac equation also has
an $so(4,2)$ invariance algebra \cite{bracken}.  
The solution spaces of these two equations each carry  hermitian 
representations of that Lie algebra.  However,  only in the massless case can this algebra
be   interpreted as the
Lie algebra of the conformal group. In the massive case, the $so(4,2)$ invariance 
algebra contains the physical
homogeneous Lorentz subalgebra, but not the generators of the physical translations, dilatations or
special conformal transformations.  

\vs\ni
What was done in Ref.\cite{bracken} 
was to construct, from the operators spanning the usual  hermitian representation
of the Poincar\'e Lie algebra on the positive-energy
solution space of the
{\it massive} Dirac equation, a set of operators that span another,  inequivalent hermitian
irreducible representation of the
Poincar\'e Lie algebra -- and of its extension, the  Lie algebra
$so(4,2)$
-- of the type appropriate for the description
of a {\it massless particle} with a definite helicity.    
What will be shown below is that if two types of four-component neutrinos are considered, and  if
a parameter with
the dimensions of mass is introduced, then we can construct from the algebra of operators for the massless
particles, 
a set of operators that span 
a representation of the Poincar\'e Lie algebra 
appropriate to the description of a spin 1/2 particle with nonzero rest mass.
In short, the operator algebra for the set of massless particles can be deformed into the operator
algebra of a
massive particle, acting on the same vector space. 

\vs\ni
The observed mixing of the three known neutrino types is thought
to be inexplicable unless some types have nonzero rest masses \cite{neutrinos}.  
Neutrino masses have not been measured directly, but it is believed that if
they are nonzero, then they are very small compared with the rest mass of the electron.
This gives renewed relevance to the
the notions of conformal invariance
in the massless case, and $so(4,2)$ invariance in the massive case, and the relationship between the
two.

\section{Massless and massive Dirac equations, and $so(4,2)$ invariance. }
 \vs\ni
 The space of positive energy solutions of the massless Dirac equation 
 \begin{equation}
 (\gamma\cdot P)\,\psi(x)\equiv \gamma_{\mu}P^{\mu}\,\psi(x)=0\,,\qquad P_{\mu}=i\partial/\partial x^{\mu}\,,
 \label{masslessdirac}
 \end{equation}
 when restricted by the helicity condition
 \begin{equation}
 \gamma_5\,\psi (x)=\pm\,\psi(x)\,,
 \label{helicity}
 \end{equation}
 carries a unitary representation of the Poincar\'e group appropriate to the description of a massless
 particle with helicity $\mp 1/2$.  Here $x=(x^{\mu})$, $\mu=0,\,1,\,2,\,3$, and the Dirac matrices
 $\gamma^{\mu}$
satisfy
 \begin{equation}
 \{\gamma^{\mu},\gamma^{\nu}\}=2g^{\mu\nu}\,,\quad \gamma_5=i\gamma_0\gamma_1\gamma_2\gamma_3\,.
 \label{diracmatrices}
 \end{equation}
We choose the diagonal metric tensor with $g^{00}=-g^{11}=-g^{22}=-g^{33}=1$, and set both $\hbar$ and
 $c$  equal to $1$.
 \vs\ni
 This
 representation of the Poincar\'e group
 extends to a unitary representation of the conformal group, with 
 generators $M_{AB}=-M_{BA}$, $A,B=0,\,1,\,2,\,3,\,5,\,6,\,$ given by \cite{barutbrittin,macksalam}

\begin{eqnarray}
M_{\mu\nu}=J_{\mu\nu}=x_{\mu}P_{\nu}-x_{\nu}P_{\mu}+\frac{1}{4}i[\gamma_{\mu},\gamma_{\nu}]\,,
\nonumber\\
\nonumber\\
M_{56}=D=x\cdot P+\frac{3}{2}i\,,\quad
M_{\mu 6}-M_{\mu 5}=P_{\mu}\,,
\nonumber\\
\nonumber\\
M_{\mu 6}+M_{\mu 5}=K_{\mu}=x_{\mu}(2D+i)-(x\cdot x)\,P_{\mu}-i(\gamma\cdot x)\gamma_{\mu} \,.
\label{Lgenerators}
\end{eqnarray}
These satisfy the commutation relations 
\begin{eqnarray}
[M_{AB},M_{CD}]=
\qquad\qquad\qquad\qquad\qquad\qquad\qquad
\nonumber\\
\nonumber\\
-i\left(
g_{AC}M_{BD}+g_{BD}M_{AC}-g_{AD}M_{BC}-g_{BC}M_{AD}\right)\,, 
\label{commrelationsA}
\end{eqnarray}
where the extended metric tensor is diagonal with $g_{55}=-1$, $g_{66}=1$. It is at once clear from
(\ref{commrelationsA}) and the form of this diagonal metric 
that the Lie algebra of the conformal group is isomorphic to $so(4,2)$.

\vs\ni
We consider a 4-component massless particle,   described by 
(\ref{masslessdirac}) but now without the fixed helicity condition (\ref{helicity}).
Noting that the generators (\ref{Lgenerators}) are not all dimensionless,
we also introduce a parameter $\mathbb M>0$ with the dimensions of a mass (or an inverse length, since
we have set $\hbar=c=1$).  Then we can
modify the
conformal algebra (\ref{Lgenerators}) 
to give a new set of dimensionless operators $N_{AB}$ that satisfy the same $so(4,2)$ 
relations (\ref{commrelationsA}) as the
$M_{AB}$, namely

\begin{eqnarray}
N_{\mu\nu}=M_{\mu\nu}=J_{\mu\nu}\,,
\quad
N_{56}=\gamma_5 D\,,\quad
\nonumber\\
\nonumber\\
N_{\mu 5}=\frac{1}{2}  \big(\mathbb M K_{\mu} - P_{\mu}/\mathbb M\big)\,,
\quad 
N_{\mu 6}=\frac{1}{2} \gamma_5 \big(\mathbb M K_{\mu} + P_{\mu}/\mathbb M\big)\,.
\label{Mgenerators}
\end{eqnarray}

\vs\ni
Our motivation for the change from the $M_{AB}$ to the $N_{AB}$,
as will become clearer below, 
is to make the operator structure in the massless case mirror that for the massive Dirac
equation 
 \begin{equation}
 (\gamma\cdot P)\,\psi(x)=m\,\psi(x)
\,,
 \label{massivedirac}
 \end{equation}
for which 
the corresponding
$so(4,2)$ invariance algebra is spanned by the operators $T_{AB}$ defined by \cite{bracken}
\begin{eqnarray}
T_{\mu\nu}= J_{\mu\nu}\,,\quad 
T_{56}= J= \frac{1}{4}\epsilon^{\mu\nu\rho\sigma}J_{\mu\nu}J_{\rho\sigma}\,,
\nonumber
\\
\nonumber\\
T_{\mu 5}=\frac{1}{2m}\,
\{J_{\mu\nu},P^{\nu}\}\,,\quad
T_{\mu 6}= \frac{1}{2m}\,\{ P_{\mu},J\}\,.
\label{Tgenerators}
\end{eqnarray} 
Here 
$\{\,,\,\}$ denotes the anticommutator, and 
$\epsilon^{\mu\nu\rho\sigma}$ is the alternating tensor with $\epsilon^{0123}=1$. 
Once again, the operators $T_{AB}$ satisfy relations of the form (\ref{commrelationsA}).

\vs\ni
The key to understanding the rather surpising $so(4,2)$ invariance of the massive Dirac equation
is that, when regarded
as {\it reducible} 
representations of the {\it homogeneous} Lorentz Lie algebra $so(3,1)$, 
two  
irreducible representations of the Poincar\'e Lie algebra are equivalent\cite{joos}:  the one
appropriate to the description of a positive
energy massless
particle with definite helicity, equal to either $+1/2$ or $-1/2$;
and the other appropriate to the description of 
a positive energy massive particle with spin $1/2$.
Each of these reducible representations of $so(3,1)$ extends to an hermitian
representation of $so(4,2)$.   
It is for this reason that  the
$J_{\mu\nu}$,  as  
generators of homogeneous Lorentz transformations for the massless particle, can be identified with
the $N_{\mu\nu}$ as in (\ref{Mgenerators}); and similarly
that as generators of
homogeneous Lorentz transformations for the massive particle, 
they can be identified with the $T_{\mu\nu}$ as in (\ref{Tgenerators}).  

\vs\ni
It is  known 
that the operators (\ref{Lgenerators})
satisfy certain characteristic `representation relations' \cite{barutbohm}, 
and the same is true of the operators (\ref{Mgenerators})
and (\ref{Tgenerators}).  
More generally, 
and more importantly,
the similar structures of the algebras of operators associated with the massless particles and with the massive
particle enable us now to proceed towards our goal.

\section{Deformation of the massless algebra to the massive algebra.}

\vs\ni
Our problem is to construct, on the space of solutions of the 4-component massless 
Dirac equation (\ref{masslessdirac}),
a set of operators $\mathbb P_{\mu}$, $\mathbb J_{\mu\nu}$ that satisfy the defining
relations of a representation of the Poincar\'e Lie algebra, appropriate for the description of a spin 1/2
particle
with rest mass $\mathbb M$.  
We do not deform the homogeneous Lorentz subalgebra, and take 
$\mathbb J_{\mu\nu}=J_{\mu\nu}$ as in (\ref{Lgenerators}).

\vs\ni
In order to proceed with the construction of $\mathbb P_{\mu}$, we first see how the operator $P_{\mu}$ is
embedded in the operator algebra associated with the massive Dirac equation 
(\ref{massivedirac}, \ref{Tgenerators}).
Although $T_{\mu 5}$ and $T_{\mu 6}$ are expressed in terms of the $P_{\mu}$ and $J_{\mu\nu}$ in
(\ref{Tgenerators}), it 
is not possible to invert these relations and express $P_{\mu}$ in terms of the $J_{\mu\nu}$,
$T_{\mu 5}$ and $T_{\mu 6}$ alone.  There are three
independent 
4-vectors acting on the space of solutions to (\ref{massivedirac}).  They could be taken to be $P_{\mu}$,
$J_{\mu\nu}P^{\nu}$ and $J_{\mu\nu}J^{\nu\rho}P_{\rho}$ but
a more convenient set for our purposes is 
\begin{eqnarray}
T_{\mu 6}+T_{\mu 5} &=&\frac{1}{2m}\{J,P_{\mu}\} +\frac{1}{2m}\{J_{\mu\nu},P^{\nu}\}\,,
\nonumber\\
\nonumber\\
&=&\frac{1}{m}\big( (J-3i/2)P_{\mu} +i W_{\mu} +J_{\mu\nu}P^{\nu}\big)\,,
\nonumber\\
\nonumber\\
T_{\mu 6}-T_{\mu 5}&=&\frac{1}{2m}\{J,P_{\mu}\} -\frac{1}{2m}\{J_{\mu\nu},P^{\nu}\}\,,
\nonumber\\
\nonumber\\
&=&\frac{1}{m}\big( (J+3i/2)P_{\mu} +i W_{\mu} +J_{\mu\nu}P^{\nu}\big)\,,
\nonumber\\
\nonumber\\
Z_{\mu}&=&\frac{1}{m}\left( -iJW_{\mu}-\frac{i}{2}J_{\mu\nu}P^{\nu} -\frac{1}{2} P_{\mu}\right)\,,
\label{RLZdef}
\end{eqnarray}
where 
$W_{\mu}=(1/2)\epsilon_{\mu\nu\rho\sigma}J^{\nu\rho}P^{\sigma}$
is the Pauli-Lubanski 4-vector, 
and $J=T_{56}$ is as in (\ref{Tgenerators}). 

\vs\ni
The first two of these 4-vector operators satisfy

\begin{eqnarray}
J\,(T_{\mu 6} + T_{\mu 5})= (T_{\mu 6} +T_{\mu 5})\,(J+i)\,,
\nonumber\\
\nonumber\\
J\,(T_{\mu 6}-T_{\mu 5})\,
=(T_{\mu 6}-T_{\mu 5})\,(J-i)\,, 
\label{massiveshiftsA}
\end{eqnarray}
which are among the relations of the form (\ref{commrelationsA}) satisfied by the $T_{AB}$, and it is
straightforward to check that, in addition,  
\begin{equation}
J\,Z_{\mu}=-Z_{\mu}\,J\,.
\label{massiveshiftsB}
\end{equation}
\vs\ni
Operators satisfying similar shifting relations
to $J$, $T_{\mu 6}+T_{\mu 5}$ and $T_{\mu 6}-T_{\mu 5}$ are easily found in the massless case, namely
$N_{56}=\gamma_5 D$, $N_{\mu 6}+N_{\mu 5}$ and $N_{\mu 6}-N_{\mu 5}$ with, 
again as a consequence of the $so(4,2)$
commutation relations of the form (\ref{commrelationsA}) satisfied by the $N_{AB}$,

\begin{eqnarray}
(\gamma_5 D) \,
(N_{\mu 6} +N_{\mu 5}) 
= 
(N_{\mu 6} +N_{\mu 5})\, (\gamma_5 D+i)\,,
\nonumber\\
\nonumber\\
(\gamma_5 D) \,
(N_{\mu 6} -N_{\mu 5}) 
= 
(N_{\mu 6} -N_{\mu 5})\, (\gamma_5 D -i)\,.
\label{masslessshiftsA}
\end{eqnarray}
To find an analogue of $Z_{\mu}$, we note that 
on solutions of (\ref{massivedirac}), 
\begin{equation}
Z_{\mu}=-\frac{i}{2}[(\gamma\cdot x)P_{\mu}-(D-i/2)\gamma_{\mu}-m(\gamma\cdot x)\gamma_{\mu}
+m x_{\mu}]\,,
\label{Zexpand}
\end{equation}
where $D$ is as in
(\ref{Lgenerators}).
This suggest the choice 
\begin{equation}
\zeta _{\mu}=
-\frac{i}{2} \tau_2 [(\gamma \cdot x) P_{\mu} - (D-i/2)\gamma_{\mu}]\,,
\label{zetadef}
\end{equation}
and it is then easily checked that indeed
\begin{equation}
(\gamma_5 D)\,\zeta_{\mu}=-\zeta_{\mu} \, (\gamma_5 D)\,,
\label{masslessshiftsB}
\end{equation} 
analogous to equation (\ref{massiveshiftsB}).  
Furthermore, it is also easily checked that 
\begin{equation}
(\gamma\cdot P)\zeta_{\mu} = - \zeta_{\mu}(\gamma\cdot P)\,,
\label{zetadirac}
\end{equation}
so that $\zeta_{\mu}$  
leaves invariant the space of solutions of the
massless Dirac equation (\ref{masslessdirac}), as do the $N_{AB}$. 
Note however that $\zeta_{\mu}$ anticommutes with $\gamma_5$.  
It is for this reason that we have  dropped
the condition (\ref{helicity}). 

\vs\ni
The relations (\ref{RLZdef}) can be inverted to give,
on solutions of the massive Dirac equation (\ref{massivedirac}),
\begin{eqnarray}
P_{\mu}= \frac{m}{2} [J^2+\frac{1}{4}]^{-1}
\qquad\qquad\qquad\qquad\qquad\qquad\qquad\qquad\qquad
\nonumber\\
\nonumber\\
\times \big( (J+i/2) (T_{\mu 6}+T_{\mu 5})
+ (J-i/2) (T_{\mu 6} -T_{\mu 5})+ 2 Z_{\mu}\big)\,.
\label{Pformula}
\end{eqnarray}
This suggests a formula for $\mathbb P _{\mu}$ on solutions of the massless Dirac equation
(\ref{masslessdirac}), namely
\begin{eqnarray}
\mathbb P_{\mu}=
\frac{\mathbb M}{2} ( D^2+\frac{1}{4})^{-1}
\qquad\qquad\qquad\qquad\qquad\qquad\qquad\qquad\qquad
\nonumber\\
\nonumber\\
\times \big( (\gamma_5 D+i/2) (N_{\mu 6}+N_{\mu 5}) 
+(\gamma_5
D -i/2)(N_{\mu 6}-N_{\mu
5}) +2 \zeta_{\mu} \big) 
\label{bbPformulaA}
\end{eqnarray}
which simplifies to 

\begin{eqnarray}
\mathbb P_{\mu}=
\frac{1}{2}(D^2+\frac{1}{4})^{-1}
\qquad\qquad\qquad\qquad\qquad\qquad\qquad\qquad\qquad
\nonumber\\
\nonumber\\
\times \big( \mathbb M^2 (D+i/2)  K_{\mu} +(D-i/2) 
P_{\mu} 
+2\mathbb M  \zeta_{\mu} \big)\,.  
\label{bbPformula}
\end{eqnarray}
Note that $[J^2 + 1/4]^{-1}$ and $[D^2+ 1/4]^{-1}$ are well-defined, nonsingular operators, and that
$\mathbb P_{\mu}$ has the appropriate dimensions. 

\vs\ni
It is now possible to check by direct calculation that the operators $\mathbb P_{\mu}$ and $\mathbb
J_{\mu\nu}=J_{\mu\nu}$ satisfy all the algebraic relations appropriate for the description of a spin 1/2
massive particle
of rest mass $\mathbb M$. 

\vs\ni
To check that 
\begin{equation}
[\mathbb P_{\mu},\mathbb P_{\nu}]=0\,, 
\label{Pcommutator}
\end{equation}
we introduce the notation
\begin{equation}
A_{\mu}B_{\nu}-A_{\nu}B_{\mu}=A_{[\mu}B_{\nu]}\,,
\label{notation}
\end{equation}
so that in particular $[\mathbb P_{\mu},\mathbb P_{\nu}]=\mathbb P_{[\mu}\mathbb P_{\nu]}$.
Then we find that, on the solution space of (\ref{masslessdirac}),  
\begin{eqnarray}
K_{[\mu}P_{\nu]}=2(D-i/2)L_{\mu\nu}-iR_{\mu\nu}\,,
\nonumber\\
\nonumber\\
P_{[\mu}K_{\nu]} =-2(D+3i/2)L_{\mu\nu}+iR_{\mu\nu}-4iS_{\mu\nu}\,,
\nonumber\\
\nonumber\\
\zeta_{[\mu}\zeta_{\nu]}
=
-2(D-i/2)L_{\mu\nu}
+2DR_{\mu\nu}
-4i(D-i/2)^2 S_{\mu\nu}\,,
\label{antirelationsA}
\end{eqnarray}
where
\begin{eqnarray}
L_{\mu\nu}=x_{\mu}P_{\nu}-x_{\nu}P_{\mu}\,,
\nonumber\\
\nonumber\\
R_{\mu\nu}= (\gamma\cdot x)(\gamma_{\mu}P_{\nu}-\gamma_{\nu}P_{\mu})\,,
\nonumber\\
\nonumber\\
S_{\mu\nu}=(i/4)[\gamma_{\mu},\gamma_{\nu}]\,.
\label{combinationsA}
\end{eqnarray}
Furthermore, again on the solutions of (\ref{masslessdirac}), we find that

\begin{eqnarray}
(D+i/2)K_{[\mu}\zeta_{\nu]}=-(D-3i/2))\zeta_{[\mu}K_{\nu]}\,,
\nonumber\\
\nonumber\\
(D-i/2)P_{[\mu}\zeta_{\nu]}=-(D+3i/2)\zeta_{[\mu}P_{\nu]}\,.
\label{antirelationsB}
\end{eqnarray}
The relations (\ref{antirelationsA}) and (\ref{antirelationsB}), together with (\ref{bbPformula}) and the
shifting formulas (\ref{masslessshiftsA}, \ref{masslessshiftsB}), 
lead to the desired
result (\ref{Pcommutator}).

\vs\ni
Similarly, to check that (\ref{bbPformula}) leads to 
\begin{equation}
\mathbb P\cdot\mathbb P=\mathbb M ^2\,,
\label{masscondition}
\end{equation}
we use the results on solutions of (\ref{masslessdirac}) 
\begin{eqnarray}
K\cdot P=2(D-i/2)(D-3i/2)\,,
\nonumber\\
P\cdot K=2(D+i/2)(D+3i/2)\,,
\label{relationsC}
\end{eqnarray}
and 
\begin{equation}
P\cdot \zeta=\zeta \cdot P=0\,,\quad K\cdot \zeta =\zeta\cdot K=0\,,\quad
\zeta\cdot \zeta =2D^2+1/2\,. 
\label{relationsD}
\end{equation}

\vs\ni
Several other useful formulas hold on the solutions of (\ref{masslessdirac}), namely

\begin{eqnarray}
\epsilon^{\mu\nu\rho\sigma}J_{\nu\rho}
K_{\sigma} = -\gamma_5 K^{\mu}\,,& &\quad
\epsilon^{\mu\nu\rho\sigma}J_{\nu\rho}
P_{\sigma} = \gamma_5 P^{\mu}\,,
\nonumber\\
\nonumber\\
\epsilon^{\mu\nu\rho\sigma}J_{\nu\rho}
\zeta_{\sigma}& =& 2i\gamma_5 D \zeta^{\mu}\,,
\label{relationsE}
\end{eqnarray}

\begin{eqnarray}
\{\gamma_5 D, K_{\mu}\}=
2\gamma_5(D-i/2) K_{\mu}\,,& &\quad
\{\gamma_5 D, \zeta_{\mu}\}= 0\,,
\nonumber\\
\nonumber\\
\{\gamma_5 D, P_{\mu}\}&=&
2\gamma_5(D+i/2) P_{\mu}\,,
\label{relationsF}
\end{eqnarray}
and

\begin{eqnarray}
\{J_{\mu\nu},K^{\nu}\}=
2(D-i/2)K_{\mu}\,,&&\quad
\{J_{\mu\nu},\zeta^{\nu}\}=0\,.
\nonumber\\
\nonumber\\
\{J_{\mu\nu},P^{\nu}\}&=&
2(D+i/2)P_{\mu}\,,
\label{relationsG}
\end{eqnarray}

\vs\ni
To obtain  the Pauli-Lubanski 4-vector associated with $\mathbb P_{\mu}$ and $\mathbb J_{\mu\nu}$, 
we use the results (\ref{relationsE}) to deduce 
from (\ref{bbPformula}) that 
\begin{eqnarray}
\mathbb W^{\mu}&=&
\frac{1}{2}\epsilon^{\mu\nu\rho\sigma}\mathbb J_{\nu\rho}\mathbb P_{\sigma}
=
\frac{1}{2}\epsilon^{\mu\nu\rho\sigma}J_{\nu\rho}\mathbb P_{\sigma}
\nonumber\\
\nonumber\\
&=& -\frac{1}{4} (D^2+1/4)^{-1}\gamma_5\left( \mathbb M^2(D+i/2)   K^{\mu} \right.
\nonumber\\
\nonumber\\
&&-\left. (D-i/2)  P^{\mu} -4i\mathbb M D
\zeta^{\mu}\right)\,.
\label{Wformula}
\end{eqnarray}
Then with the help of (\ref{relationsC}, \ref{relationsD}) we get 
\begin{equation}
\mathbb W\cdot \mathbb W =-\frac{1}{2}(\frac{1}{2}+1)\mathbb M^2\,,
\label{spincondition}
\end{equation}
as required for a massive particle with spin $1/2$.  

\vs\ni
Finally, we note with the help of (\ref{relationsF}, \ref{relationsG})
that in terms of $\mathbb P_{\mu}$ and $\mathbb J_{\mu\nu}$, the $so(4,2)$
generators 
(\ref{Mgenerators}) 
in the massless
case 
can be rewritten as

\begin{eqnarray}
N_{\mu\nu}= \mathbb J_{\mu\nu}\,,\quad 
N_{56}= \gamma_5 D =\frac{1}{4}\epsilon^{\mu\nu\rho\sigma}\mathbb J_{\mu\nu}\mathbb J_{\rho\sigma}\,,
\nonumber
\\
\nonumber\\
N_{\mu 5}=\frac{1}{2\mathbb M}\,
\{\mathbb J_{\mu\nu},\mathbb P^{\nu}\}\,,\quad
N_{\mu 6}= \frac{1}{2\mathbb M}\,\{ \mathbb P_{\mu},\gamma_5 D\}\,.
\label{newMgenerators}
\end{eqnarray} 
which have the same form as the formulas (\ref{Tgenerators}) in the massive case.

\section{Concluding remarks.}
The algebra of operators associated with a 4-component massless particle has been deformed 
to obtain the operator algebra associated with a spin 1/2 particle with positive rest mass. 

Instead of dropping the helicity condition (\ref{helicity}), we could instead have introduced two flavors of
2-component massless particles, and an associated set of Pauli flavor matrices $\tau_1$, $\tau_2$, $\tau_3$
that commute with all Dirac matrices.  Then we could have replaced $\gamma_5$ in 
(\ref{helicity}) by $\tau_3\gamma_5$, and our calculations would go through as before provided that we
replaced $\zeta_{\mu}$  everywhere, and in particular in (\ref{bbPformula}) and (\ref{Wformula}),
by  $\tau_1 \zeta_{\mu}$.
The operators $\mathbb P_{\mu}$ and $\mathbb W_{\mu}$ so modified leave invariant the space of solutions of
(\ref{masslessdirac}) and the modified helicity condition.   
In this way, the operator algebra associated with 
two flavors of 2-component massless particles can also be deformed to obtain
the algebra of a single spin 1/2 particle with positive rest mass.  

From a geometric
point of view it is notable that non-null 4-vector operators $\mathbb P_{\mu}$ and $\mathbb W_{\mu}$
can be constructed on the state space of a set of massless particles, for which the 4-vector 
operators $P_{\mu}$ and
$K_{\mu}$ are null.  

It is  hoped
that
these results can be exploited to describe the mixing of neutrino types without
the introduction of neutrino rest masses.


\begin{thebibliography}{99}

\bibitem{barutbrittin}
Barut, A.O. and Brittin, W.E., {\it De Sitter and Conformal Groups and their Applications},
Lecture Notes in Theoretical Physics Vol. XIII (Colorado Associated University Press, Boulder, 1971).

\bibitem{canterbury}
Barut, A.O., {\it Dynamical Groups and Generalized Symmetries in Quantum Theory},
(University of Canterbury, Christchurch, NZ, 1971).  

\bibitem{barutbracken}
Barut, A.O. and Bracken, A.J., ``Zitterbewegung and the internal geometry of the electron," 
{\it Phys. Rev. D} {\bf 23} (1981), 2454--2463.  

\bibitem{macksalam}
Mack, G. and Salam, A., ``Finite-component field representations of the conformal group,"
{\it Ann. Phys.} {\bf 53} (1969), 174--202.

\bibitem{bracken}
Bracken, A.J., ``$O(4,2)$: an exact invariance algebra for the electron,"
{\it J. Phys. A} {\bf 8} (1975), 808--815.  

\bibitem{neutrinos}
Conrad, J.M., ``A review of experimental results on neutrino oscillations," {\it Nucl. Phys. A} {\bf 663}
(2000), 210C--271C. 

\bibitem{macktodorov}
Mack, G. and Todorov, I.T., ``Irreducibility of the ladder representations of $U(2,2)$ when
restricted to the Poincar\'e group," {\it J. Math. Phys. }{\bf 10} (1969), 2078--2085. 

\bibitem{joos}
Joos, H., ``Theory of representations of the inhomogeneous Lorentz group as a basis for quantum
mechanical kinematics," {\it Forts. der Physik} {\bf 10} (1962), 65-146. 

\bibitem{barutbohm}
Barut, A.O. and Bohm, A., ``Reduction of a class of $O(4,2)$ representations with respect to $SO(4,1)$
and $SO(3,2)$," {\it J. Math. Phys. } {\bf 11} (1970), 2938--2945.

\end{thebibliography}
\end{document}